\documentclass[dvips]{article}
\usepackage{icrctc07}

\title{Upper Limits from HESS Observations of AGN in 2005-2007}

\shorttitle{Upper Limits from HESS Observations of AGN}
\authors{W.\,Benbow$^{1}$ and R.\,B\"uhler$^{1}$ for the HESS Collaboration}
\shortauthors{W.\,Benbow et al.}
\afiliations{$^1$ Max-Planck-Institut f\"ur Kernphysik, Heidelberg, Germany}
\email{Wystan.Benbow@mpi-hd.mpg.de}

\abstract{
Very high energy (VHE; $>$100 GeV) observations of a sample of 
selected active galactic nuclei (AGN) were performed
between January 2005 and April 2007 with the High Energy Stereoscopic System (HESS),
an array of imaging atmospheric-Cherenkov telescopes.
Significant detections are reported elsewhere for many of these objects.  
Here, integral flux upper limits for twelve candidate
very-high-energy (VHE; $>$100 GeV)  gamma-ray emitters are presented.
In addition, results from HESS observations of four known VHE-bright AGN
are given although no significant signal is measured.  For three of
these AGN (1ES\,1101$-$232, 1ES\,1218+304, and Mkn\,501) simultaneous 
data were taken with the Suzaku X-ray satellite.}

\begin{document}
\maketitle

\section{Introduction}

The HESS array \cite{HESS_jim} of four imaging atmospheric-Cherenkov 
telescopes located in Namibia is used to search for
VHE $\gamma$-ray emission from various
classes of astrophysical objects.
Approximately 300 hours per year, $\sim$30\% of the total
observation budget, are dedicated to
HESS studies of AGN. These observations are
divided between monitoring the flux of
known VHE-bright AGN and searching for new VHE sources. 
For the monitoring observations, an 
AGN is typically observed for a few hours, distributed
over several nights, a month for $\sim$3 months, with
the hopes of detecting a bright flaring episode (see, e.g., \cite{2155_flare}).
In the discovery part of the AGN program, a candidate from 
a large, diverse sample of relatively nearby AGN is typically
observed for $\sim$10 hours.  If any of these observations 
show an indication for a signal (e.g., an excess with significance
more than $\sim$3 standard deviations), a deeper 
exposure is promptly scheduled to increase the overall significance
of the detection and to allow for a spectral measurement.

The targets of HESS AGN observations are primarily blazars, 
a class which includes both BL\,Lac objects and
Flat Spectrum Radio Quasars (FSRQ). The spectral energy
distributions (SEDs) of these objects are generally
characterized by two peaks: a lower-energy one in the optical to X-ray
regime, and another which potentially extends to
$\gamma$-ray energies. Based on their SEDs, 
BL\,Lacs are generally categorized into groups that are either 
low (LBL), intermediate (IBL), or high-frequency-peaked (HBL).
An overwhelming majority of VHE-emitting AGN
are HBL, therefore these objects
are the primary targets of HESS AGN observation
program.  However, prominent examples of different
types of AGN are also observed with the hopes of 
detecting new AGN classes.  These include
radio-loud objects such as Fanaroff-Riley (FR) galaxies
and narrow line Seyfert (NLS) galaxies, and radio-weak objects
like typical Seyfert (Sy) galaxies,  all of which come
in several types (generally I or II).

For all the following results, the HESS standard analysis \cite{std_analysis} is used.
All upper limits are given at the 99\% confidence level \cite{UL_tech}.
The flux quantities are calculated assuming a power-law spectrum
with photon index $\Gamma$=3.0, with the exception of those for 1ES\,1101$-$232
where  $\Gamma$=2.94, as measured \cite{Gerd_1101} in 2004-05, is chosen.
The reported values change by less than $\sim$10\% when a 
different photon index (i.e. $\Gamma$ between 2.5 and 3.5) 
is assumed.  The effects of changes in the absolute optical efficiency
of HESS are corrected for using efficiencies 
determined from simulated and observed muons \cite{HESS_crab}. 
The systematic error on all flux quantities is estimated to
be $\sim$20\%.

\section{Limits from Discovery Observations}

   \begin{table*}[ht]
      \caption{The candidate AGN in groups of blazars and non-blazars. 
The asterisk denotes the four candidates detected by the EGRET satellite 
\cite{EGRET_catalog}. The redshift ($z$), total good-quality
live time (T), mean zenith angle of observation (Z$_{\mathrm{obs}}$),
the observed excess and significance (S) are shown.
Integral flux upper limits above the energy threshold of the observations (E$_{\mathrm{th}}$), 
and the corresponding percentage of the HESS Crab Nebula flux \cite{HESS_crab} above the same
threshold, are also shown.  The flux units are $10^{-12}$ cm$^{-2}$ s$^{-1}$. The $\dagger$
represents the six upper limits which are the most constraining ever reported for the object.}
         \label{results} 
        \centering
         \begin{tabular}{c c c c c c c c c c}
            \hline\hline
            \noalign{\smallskip}
	     Object & $z$ & Type & T & Z$_{\mathrm{obs}}$ & Excess & S & E$_{\mathrm{th}}$ & I($>$E$_{\mathrm{th}}$) & Crab \\
             & &  & [hrs] & [$^{\circ}$] & & [$\sigma$] & [GeV] & [f.u.] & \%  \\
            \noalign{\smallskip}
            \hline
            \noalign{\smallskip}
	{\it Blazar} \\
            \noalign{\smallskip}
             III\,Zw\,2           & 0.0893 & FSRQ & 1.7 & 37 & 12 & 1.4 & 420 & 5.36$^{\dagger}$ & 6.4 \\ 
	     BWE\,0210+116$^{*}$  & 0.250  & LBL & 6.0 & 43 & $-$13 & $-$0.9 & 530 & 0.72$^{\dagger}$ & 1.2\\
	     1ES\,0323+022        & 0.147  & HBL & 7.2 & 27 & 13 & 0.7 & 300 & 2.52 & 1.9\\
	     PKS 0521$-$365$^{*}$ & 0.0553 & LBL & 3.1 & 26 & 11 & 0.8 & 310 & 5.40$^{\dagger}$ & 4.2\\
	     3C\,279$^{*}$	  & 0.536 & FSRQ & 2.0 & 26 & 5 & 0.5 & 300 & 3.98$^{\dagger}$ & 2.9\\
             RBS\,1888            & 0.226 & HBL	& 2.4 & 15 & 30 & 2.2 & 240 & 9.26 & 4.9\\
             PKS\,2316$-$423      & 0.055 & IBL	& 4.1 & 20 & 29 & 1.6 & 270 & 4.74 & 3.0\\
 	     1ES\,2343$-$151      & 0.226 & IBL	& 8.6 & 17 & $-$16 & $-$0.6 & 230 & 2.45$^{\dagger}$ & 1.2\\
\\	{\it Non-blazar} \\
            \noalign{\smallskip}
             NGC\,1068		& 0.00379 & Sy II & 1.8 & 29 & 9 & 1.1 & 330 & 5.76 & 4.9\\
             Pictor\,A		& 0.0342 & FR II & 7.9 & 31 & $-$23 & $-$1.1 & 320 & 2.45 & 2.0\\
	     PKS\,0558$-$504	& 0.137 & NLS I & 8.3 & 28 & $-$14 & $-$0.7 & 310 & 2.38$^{\dagger}$ & 1.8\\
             NGC\,7469		& 0.0164 & Sy I	& 3.4 & 34 & $-$14 & $-$1.3 & 330 & 1.38 & 1.2\\
          \noalign{\smallskip}
            \hline
       \end{tabular}
   \end{table*}

Twenty-nine AGN were observed by HESS from January 2005 through April 2007.
Some of these objects were previously shown by HESS
to emit VHE $\gamma$-rays, and the discoveries of VHE emission
from others are reported elsewhere. Of the remaining AGN
with non-zero good-quality exposure,
twelve show no indication of any VHE emission.  
As many of the HBL observed by HESS have been detected, 
the twelve candidates discussed in this section are largely not HBL.
Table~\ref{results} shows these AGN, their
redshift and AGN type, as well as details of their observations.
The mean good-quality exposure for 
the candidates is 4.7 hours live time at a
mean zenith angle of 28$^{\circ}$. In 5 hours of observations, 
the sensitivity of HESS \cite{std_analysis}
enables a 5$\sigma$ detection of an $\sim$2\% Crab Nebula flux 
source at 20$^{\circ}$ zenith angle.

\begin{figure}
\begin{center}
\noindent
\includegraphics [width=0.5\textwidth]{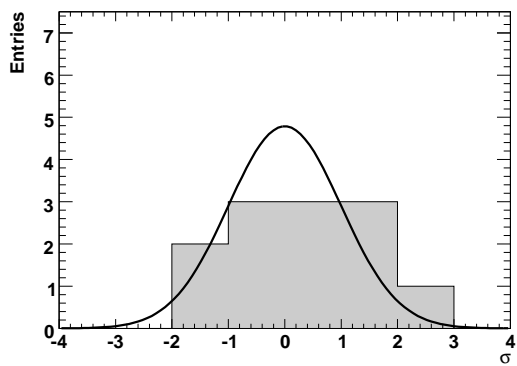}
\end{center}
      \caption{Distribution of the significance observed 
from the twelve candidate AGN.  The curve
represents a Gaussian distribution with zero mean 
and a standard deviation of one.}
         \label{AGN_sigma}
\end{figure}

As mentioned previously, no significant excess of VHE $\gamma$-rays is found 
from any of these twelve AGN in the given exposure time. 
Figure~\ref{AGN_sigma} shows the distribution 
of the significance observed from the direction of each AGN.  
The measured excess, corresponding significance and resulting
integral flux limits are given in Table~\ref{results} 
for each AGN.  Six of the upper limits are the most constraining
ever reported from these objects, and 
the other six limits are only surpassed by those from
HESS observations in 2004 \cite{HESS_AGN_UL}.
Combining the excess from all twelve candidates only 
yields a total of 29 events 
and a statistical significance of
1.1$\sigma$.  No significant excess is found in a search for serendipitous 
source discoveries in the HESS field-of-view centered on each of the 
AGN.  Further, as the nightly flux from each target is well-fit by a
constant, no evidence for VHE flares is found from any of the twelve AGN.

\section{Low Altitude HESS Observations}

Three northern AGN, known to emit VHE $\gamma$-rays,
were briefly (good-quality live time $<$2.2 h) 
observed at low altitudes with HESS. 
At such altitudes the threshold of HESS is higher 
and the sensitivity is reduced.  However,
observations at low altitudes sample
the VHE spectrum at much higher energies than
the typical near-zenith observations made with
Cherenkov-telescope arrays.  Simultaneous measurements
of the same northern target with HESS and a Northen Hemisphere
instrument enable both the determination of the object's spectrum
over several orders of magnitude in energy, as well as
cross-calibration between the instruments 
(see, e.g., \cite{HESS_MAGIC_421}).
For two of these targets (1ES\,1218+304 and Mkn\,501)
simultaneous observations
were successfully performed by the MAGIC VHE telescope
and the Suzaku X-ray satellite \cite{Suzaku_info}.

HESS observed Mkn\,421 on April 12, 2005.  
The good-quality exposure is 0.9 h live time at a 
mean zenith angle of $63^{\circ}$.
A marginally significant excess (28 events, 3.5$\sigma$)
is found.  The corresponding integral flux
above the 2.1 TeV analysis threshold is  
I($>$2.1 TeV) = $(3.1\pm1.0_{\rm stat}) \times 10^{-12}$
cm$^{-2}$\,s$^{-1}$, or 45\% of the HESS Crab Nebula flux above
the same threshold. 

The HESS observations of 1ES\,1218+304 on May 19, 2006
yield a good-quality data set of 1.8 h live time at a 
mean zenith angle of $56^{\circ}$.
The resulting excess is not significant (9 events, 
1.2$\sigma$). The upper limit on the 
integral flux above the 1.0 TeV analysis threshold is  
I($>$1.0 TeV) $ < 3.9 \times 10^{-12}$
cm$^{-2}$\,s$^{-1}$.  This corresponds to
17\% of the HESS Crab Nebula flux 
above the same threshold.

HESS observations of Mkn\,501 occurred on July 18, 2006.  
All data pass the standard quality-selection
criteria, yielding an exposure of 2.2 h live time at a 
mean zenith angle of $64^{\circ}$.
Mkn\,501 is not detected by HESS
as the resulting excess is $-9$ events ($-0.8$$\sigma$).
The upper limit on the integral flux above the 2.5 TeV analysis threshold is  
I($>$2.5 TeV) $ < 1.1 \times 10^{-12}$
cm$^{-2}$\,s$^{-1}$, or 22\% of the 
HESS Crab Nebula flux above the same threshold.

\section{VHE Monitoring of 1ES\,1101$-$232\label{1101_Sect}}
\begin{figure}
\begin{center}
\noindent
\includegraphics [width=0.45\textwidth]{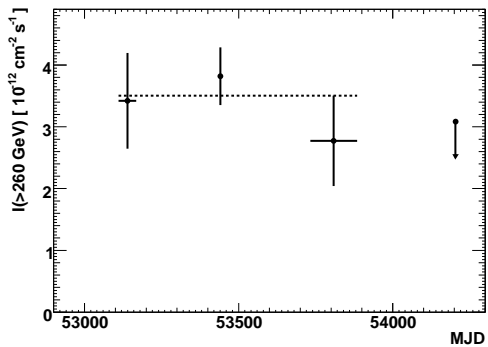}
\end{center}
      \caption{The annual light curve, I($>$260 GeV), 
from HESS measurements of 1ES\,1101$-$232. 
The upper limit in 2007 is at the 99.9\% confidence level.
The 2004 and 2005 data are published elsewhere \cite{Gerd_1101}.  
The actual observation dates are shown by the x-error bars.  
The dashed line is the average flux measured from 2004-2006. }
         \label{1101_lc}
\end{figure}

1ES\,1101$-$232 was discovered by HESS \cite{Nature_EBL,Gerd_1101} 
to emit VHE $\gamma$-rays during observations in 2004-2005.
As part of a campaign to monitor its VHE flux, it was re-observed 
for a total (good-quality observations) of 18.3 h in 2006-07.
A marginally significant excess (117 events, 3.6$\sigma$) 
is detected from 1ES\,1101$-$232
in the 2006 observations (13.7 h), and the object is not detected 
(16 events,  0.9$\sigma$) in 2007.   As can be seen from Figure~\ref{1101_lc}
the upper limit from 2007 falls below the average flux measured by
HESS from 2004-2006.  Some of the 2006 
HESS data (4.3 h) are simultaneous with Suzaku X-ray 
observations.  In these data, the blazar is again marginally
detected (51 events,  0.9$\sigma$) and the corresponding flux
is I($>$260 GeV) $ < (3.2\pm1.4_{\rm stat}) \times 10^{-12}$
cm$^{-2}$\,s$^{-1}$.

\section{Discussion \& Conclusions}

One of the defining characteristics of AGN is their extreme
variability. The VHE flux from any of these AGN may increase 
significantly during future flaring episodes 
(see, e.g., \cite{2155_flare}) and could potentially
exceed the limits presented here.  In
addition, accurate modeling of the SED 
requires that the state of the source is accounted for.  
Therefore, in the absence of contemporaneous observations
at lower energies,  it is recommended that these results 
be conservatively interpreted as limits on,
or measurements of, the steady-component or quiescent flux from the AGN.
Clearly, the simultaneous Suzaku X-ray data 
from Mkn\,501, 1ES\,1218+304, and 1ES\,1101$-$232, make the
HESS results from these objects particularly useful.
Finally, interpretation of the SED of an AGN not only requires accounting
for the state of the source, but also the redshift and energy dependent
absorption \cite{EBL_1} of VHE photons on the 
Extragalactic Background Light (EBL),
which is potentially large \cite{Nature_EBL,EBL_2} for some of these sources.

With the detection of ten VHE AGN, including the discovery of
seven, the HESS AGN observation program has been highly successful. 
However, despite more than five years of operations, the observation 
program is not complete as many proposed candidates 
have either not yet been observed
or only have a fraction of their intended exposure.  
Therefore, the prospects of finding additional VHE-bright AGN 
with HESS are still excellent.

\section{Acknowledgements}

The support of the Namibian authorities and of the University of Namibia
in facilitating the construction and operation of H.E.S.S. is gratefully
acknowledged, as is the support by the German Ministry for Education and
Research (BMBF), the Max Planck Society, the French Ministry for Research,
the CNRS-IN2P3 and the Astroparticle Interdisciplinary Programme of the
CNRS, the U.K. Science and Technology Facilities Council (STFC),
the IPNP of the Charles University, the Polish Ministry of Science and 
Higher Education, the South African Department of
Science and Technology and National Research Foundation, and by the
University of Namibia. We appreciate the excellent work of the technical
support staff in Berlin, Durham, Hamburg, Heidelberg, Palaiseau, Paris,
Saclay, and in Namibia in the construction and operation of the
equipment.


\end{document}